\documentclass[table]{ccjnl}
\usepackage{lipsum,amsmath}
\usepackage{cuted}
\usepackage{bm}

\usepackage{subfloat}
\graphicspath{{figures/}}

\title{Dynamic Write-Voltage Design and Read-Voltage Optimization for MLC NAND Flash Memory}

\author{Runbin Cai\inst{1}, Yi Fang\inst{1,*}, Zhifang Shi\inst{1}, Lin Dai\inst{1}, Guojun Han\inst{1}\corinfo{fangyi@gdut.edu.cn}}
\receiveddate{}
\reviseddate{}
\Editor{}

\address[1]{School of Information Engineering, Guangdong University of Technology, China}

\begin{document}

\maketitle

\begin{abstract}
To mitigate the impact of noise and interference on multi-level-cell (MLC) flash memory with the use of low-density parity-check (LDPC) codes, we propose a dynamic write-voltage design scheme considering the asymmetric property of raw bit error rate (RBER), which can obtain the optimal write voltage by minimizing a cost function.
In order to further improve the decoding performance of flash memory, we put forward a low-complexity entropy-based read-voltage optimization scheme, which derives the read voltages by searching for the optimal entropy value via a log-likelihood ratio (LLR)-aware cost function.
Simulation results demonstrate the superiority of our proposed dynamic write-voltage design scheme and read-voltage optimization scheme with respect to the existing counterparts.
\keywords{Multi-level-cell (MLC); NAND flash memory; write voltage; read voltage; error correction coding}
\end{abstract}

\section{Introduction}

As a non-volatile memory (NVM) device, NAND flash memory has attracted extensive attention from both academia and industry due to the advantages of high capacity and low power consumption.
Solid state disk (SSD) based on NAND flash memory has gradually replaced traditional storage devices and been widely used in various scenarios (e.g., mobile intelligent devices).
With the era of Internet of Things (IoT), the generation of a large amount of data has brought great challenges to the capacity and reliability of storage devices.
To address the above issue, a great deal of research effort has been devoted to boosting the storage capacity and reliability of flash memory.
For single-level-cell (SLC) flash memory, each cell stores one bit. With the emergence of multi-level-cell (MLC), triple-level-cell (TLC), quadruple-level-cell (QLC) and three-dimentional (3D) flash memory \cite{9264190}, the capacity of flash memory has been significantly increased.
In MLC flash memory, each cell stores two bits, which are represented as $\{11, 10, 00, 01\}$. The bit on the left is called the most significant bit (MSB), while the bit on the right is called the least significant bit (LSB).
However, the increase of storage capacity and the decrease of processing size make the flash memory suffer from more serious noises and interferences, mainly including programming noise, random telegraph noise (RTN), data retention noise and cell-to-cell interference (CCI) \cite{9169701}.
These noises and interferences greatly reduce the storage reliability of flash memory.

To enhance the robustness against noise and interference, error-correction-coding (ECC) technology is applied to MLC flash memory.
For example, Bose-Chaudhuri-Hocquenghem (BCH) codes have been adopted as the ECC scheme for flash memory in \cite{8351503}. However, the BCH codes are more suitable for short-codeword-length scenario, which is extremely difficult to meet the error-correction requirements of flash memory.
Thereby, low-density parity-check (LDPC) codes have been widely studied in flash memory because of its low implementation complexity and strong error-correction ability, especially in the long-codeword-length scenario.
Moreover, due to the influence of channel-asymmetric noise, the probability distribution functions (PDFs) of the four voltages in MLC flash memory continue changing over different program-and-erase (PE) cycles and retention time.
Consequently, it is necessary to update the write voltage and optimize the read voltage for adapting to the variation of flash memory channel.

\subsection{Literature and Motivation}
To attenuate the asymmetric noise errors of flash memory, some techniques have been presented to optimize write voltage and read voltage.
In \cite{1888888}, the authors have proposed a write-voltage optimization scheme, which only considers the RTN noise in the flash memory channel.
Furthermore, another scheme has been proposed in \cite{1999999} to optimize the write voltage by modeling the flash memory channel as an additive white Gaussian noise (AWGN) channel.
However, these techniques are not precise and suitable for flash memory channel, because the practical NAND flash memory contains various non-stationary noises such as the programming noise, RTN, data retention noise and the CCI.
These noises and interferences have different characteristics and vary continuously with the PE cycles and the retention time.

On the other hand, differential evolution algorithm has been utilized to find the optimal write voltage by maximizing the channel capacity (MCC) of flash memory channel in \cite{2000000}.
A write-voltage design scheme has been also proposed in \cite{9169701} to design the efficient write voltage by minimizing the raw bit error rate (RBER) of flash memory channel.
However, the authors have not considered the unbalanced RBERs between MSB and LSB pages.
In \cite{8075860}, the authors have considered the unbalanced RBERs and searched the write voltage to minimize the RBER difference (MRD) between MSB and LSB pages.
Nonetheless, this write-voltage scheme does not consider the influence of coding on flash memory system.
To further improve the reliability of flash memory, more noises and interferences, the unbalanced RBER property as well as LDPC codes are considered in our proposed write-voltage design scheme.

To enable soft-decision decoding of error-correction codes, NAND flash memory systems usually need the fine-grained memory-sensing operations. Nonetheless, high sensing accuracy always lead to severe sensing delay.
Therefore, it is very important to balance the decoding performance and delay in the read-voltage design of flash memory.

The conventional read-voltage design scheme has utilized the uniform quantization scheme \cite{8314735}.
This method has read the data by evenly placing the read voltage.
However, the log-likelihood-ratio (LLR) information obtained by the uniform quantization scheme is not accurate.
In \cite{8708250}, the authors have proposed a ``constant ratio (CR)''  non-uniform quantization scheme to improve the memory-sensing accuracy.

In \cite{2555555}, the authors have proposed the convolutional neural network (CNN)-based detection scheme to obtain the read voltages and LLR values.
Nevertheless, these LLR values are not accurate because the network is used when the channel model is unknown, and it uses training data to obtain the conditional probability distribution of the voltage states corresponding to each threshold voltage, while the probability distribution of the threshold voltage cannot be obtained.
The authors of \cite{7152879} have proposed the adaptive read-voltage (ART) algorithm
to use the statistical data for calculating the mean and variance of each voltage-state distribution,
and then obtain the read voltages and calculate LLR values.
But in practice, this scheme cannot obtain the accurate threshold voltages of the memory cells, and thus leads to inaccurate LLR values.

A read-voltage optimization scheme based on maximum mutual information (MMI) has been also proposed in \cite{8708250}.
However, this scheme obtains the read voltages
under the assumption that the codeword length is infinite.
The authors in \cite{9169701} have also proposed to exploit the entropy function to select the read voltage for controlling the width of the erasure area, so as to find the read voltage that minimizes the bit error rate (BER).
Nevertheless, in \cite{9169701}, the optimal entropy is obtained through the iterative decoding (i.e., belief-propagation (BP) decoding) operation, which suffers from high computational complexity.

\subsection{Our Contributions}
The principle idea behind this paper is to improve the decoding performance of LDPC-coded flash memory through optimization of write voltage and read voltage over the PE cycles and retention time.
To achieve this goal, we first propose a cost-function-aided method to update the write voltage, called {\em dynamic write-voltage design scheme}.
Furthermore, motivated by the importance of the LLR for decoding performance, we present a new read-voltage optimization scheme to adjust the position of read voltage to meet the variation of flash memory channel, which enables relatively lower complexity than the existing entropy-based read-voltage design scheme.
Simulation results illustrate that the proposed dynamic write-voltage design scheme and the read-voltage optimization scheme can effectively improve the BER performance of flash memory compared with the state-of-the-art schemes.

The rest of the paper is organized as follows. In Sect.~\ref{channel model}, we depict the channel model of MLC flash memory. In Sect.~\ref{write-voltage-design}, we propose a novel cost-function-aided dynamic write-voltage design scheme for the flash memory system. The read-voltage optimization scheme is conceived in Sect.~\ref{read-thresholds-optimization}. Simulation results are shown and discussed in Sect.~\ref{Simulation Results}. Finally,
the conclusions are drawn in Sect.~\ref{Conclusions}.

\section{Channel Model of Flash Memory}
\label{channel model}
The practical NAND flash memory contains various non-stationary noises such as programming noise, RTN, data retention noise and CCI \cite{9169701, 6484076}.
These noises and interferences have different characteristics and vary continuously over the PE cycles and retention time. Fig.~\ref{Fig1} shows the threshold-voltage distribution in an MLC flash memory with Gray mapping.
The write voltages for the four threshold-voltage states, i.e., $V_{\rm{min}}, V_1, V_2$ and $V_{\rm{max}}$, represent data symbols of $11, 10, 00, 01$, respectively.

\subsection{Programming Noise}
In an MLC flash memory, the programming noise follows a Gaussian distribution \cite{6484076}. The PDF of the programming noise is expressed by
\begin{eqnarray}\label{eq:programming noise}
p_{p_i}(x) = \left\{
\begin{aligned}
\mathcal{N}(0, \sigma_e^2)&&{\rm if} ~i = {\rm 11} \\
\mathcal{N}(0, \sigma_p^2)&&{\rm if} ~i \neq 11
\end{aligned}
\right.,
\end{eqnarray}
where $i \in \{11, 10, 00, 01\}$.
In practice, the flash memory utilizes the iterative incremental step pulse programming (ISPP) algorithm for programming operation \cite{8819688, 9416922}.
Thus, the threshold-voltage distribution of the noise-free programming cell can be modeled as a uniform distribution \cite{6484076}
\begin{eqnarray}\label{eq:uniform-distribution}
p_{v_{\rm{pp}}}(x) = \left\{
\begin{aligned}
{\frac {1}{v_{\rm pp}}},&&{\rm for}~0\leq{x}\leq{v_{\rm pp}} \\
0,&&{\rm otherwise}
\end{aligned}
\right.,
\end{eqnarray}where $v_{\rm pp}$ is the step size of programming voltage.

Afterwards, the programming-noise distribution for the programming cell
can be calculated via the convolution operation (i.e., ``$\ast$'') of uniform and Gaussian distribution functions, given by \cite{6484076}
\begin{eqnarray}\label{eq:programming_states noise}
p_{n_i}(x) = p_{p_i}(x) \ast p_{v_{\rm pp}}(x),
\end{eqnarray}
where $p_{n_i} \in \{p_{n_{10}}, p_{n_{00}}, p_{n_{01}}\}$ for $V_i \in \{V_1, V_2, V_{\rm max}\}$.

\subsection{Random Telegraph Noise}
In the NAND flash memory, PE cycling causes damage to the tunnel oxide of floating gate transistors, which directly results in threshold-voltage shift and fluctuation.
This is referred to as {\rm random telegraph noise}. In particular, the RTN is a non-stationary noise and the PDF can be defined as a Gaussian distribution \cite{9169701}:
\begin{align}\label{eq:RTN}
p_{t}(x) =  \frac{1}{\sqrt{2\pi}\sigma_t} e^{-\frac{x^2}{2\sigma_t^2}},
\end{align}
where $\sigma_t^2$ is the variance of the PDF for RTN, which is affected by the PE cycles.

\subsection{Data Retention Noise}
Data retention noise, caused by the leakage of charge from the floating gate over time, is one of the dominant errors in the flash
memory.
According to \cite{2111111}, the data retention noise is approximately considered to follow Gaussian distribution whose mean and variance are affected by the data retention time and the number of PE cycle.
The PDF is expressed by \cite{9169701,2111111}:
\begin{align}\label{eq:data retention noise}
p_{r_i}(x) =  \frac{1}{\sqrt{2\pi}\sigma_{r_i}} e^{-\frac{(x-\mu_{r_i})^2}{2\sigma_{r_i}^2}},
\end{align}
where $\mu_{r_i}$ and $\sigma_{r_i}^2$ represent the mean and variance of the PDF for data retention noise, and $i \in \{11, 10, 00, 01\}$.
Especially, the mean $\mu_{r_i}$ and variance $\sigma_{r_i}^2$ can be calculated as \cite{9169701}
\begin{align}\label{eq:DRN-expectation}
\mu_{r_i} =& [{A_r}\cdot(PE)^{{\alpha}_1}+{B_r}\cdot(PE)^{{\alpha}_0}] \nonumber \\
 &\times (V_{r_i} - x_0)\cdot\log(1+T),
\end{align}
\begin{align}
\label{eq:DRN-variance}
\sigma_{r_i} = 0.4\cdot|\mu_{r_i}|.
\end{align}

\subsection{Cell-to-Cell Interference}
The cell-to-cell interference is caused by the effect of parasitic coupling capacitance between adjacent cells.
Hence, the threshold voltage of the victim cell will increase during the programming prodecure.
According to \cite{8314735}, the voltage-shift value of the victim cell can be expressed as
\begin{align}\label{eq:CCI}
\Delta V_{\rm{CCI}} = \sum\limits_k \Delta V_k \gamma_k,
\end{align}
where $\Delta V_{\rm{CCI}}$ is the change of threshold voltage of the victim cell, $\Delta V_k$ is the shift of the threshold voltage in the $k$-th interference cell, and $\gamma_k$ is the capacitive coupling coefficient between the victim cell and the $k$-th interference cell.
As the CCI in the read-back voltage can be eliminated by employing the post-compensation technique \cite{5460923}, we ignore this interference in this work.

\subsection{The Overall threshold voltage distribution}
According to \cite{9169701}, we can approximately describe the threshold voltages by a Gaussian distribution, i.e.,
\begin{align}\label{eq:the final threshold voltage distribution}
p_{s_i}(x) =  \frac{1}{\sqrt{2\pi}\sigma_{s_i}} e^{-\frac{(x-\mu_{s_i})^2}{2\sigma_{s_i}^2}},
\end{align}
where $i \in \{11, 10, 00, 01\}$,
the means and variances of the four voltage states are as follows:
\begin{eqnarray}\label{eq:mean}
\mu_{s_{11}} =& V_{\rm{min}}-\mu_{r_{11}} , \nonumber \\
\sigma_{s_{11}} =& \sqrt{\sigma_e^2+\sigma_t^2+\sigma_{r_{11}}^2} , \nonumber \\
\mu_{s_{10}} =& V_1+{v_{\rm pp}}/2-\mu_{r_{10}} , \nonumber \\
\sigma_{s_{10}} =& \sqrt{\sigma_p^2+\sigma_t^2+\sigma_{r_{10}}^2} , \nonumber \\
\mu_{s_{00}} =& V_2+{v_{\rm pp}}/2-\mu_{r_{00}} , \nonumber \\
\sigma_{s_{00}} =& \sqrt{\sigma_p^2+\sigma_t^2+\sigma_{r_{00}}^2} , \nonumber \\
\mu_{s_{01}} =& V_{\rm{max}}+{v_{\rm pp}}/2-\mu_{r_{01}} , \nonumber \\
\sigma_{s_{01}} =& \sqrt{\sigma_p^2+\sigma_t^2+\sigma_{r_{01}}^2} . \nonumber
\end{eqnarray}
In this paper, we set the channel model parameters as the same in \cite{9169701}: $V_{\rm{min}} = 1.4, V_{\rm{max}} = 3.93, \sigma_e = 0.35, \sigma_p = 0.05, v_{\rm pp} = 0.3, \sigma_t = 0.00025{(PE)}^{0.62}, A_r = 0.000055, B_r = 0.000235, \alpha_1 = 0.62, \alpha_0 = 0.32$ and $x_0 = 1.4$. 

\begin{figure}[!tbp]
\centering
\includegraphics[width=0.5\textwidth]{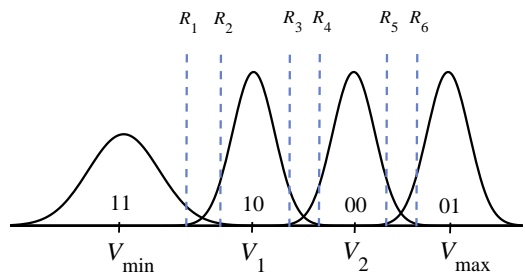}
\vspace{-0.3cm}
\caption{Distribution of threshold voltages in an MLC flash memory with $6$-level quantization scheme.}
\label{Fig1}
\end{figure}

\section{Proposed Dynamic Write-Voltage Design Scheme}\label{write-voltage-design}
To improve the error-rate performance of LDPC-coded flash memory, the write-voltage levels are required to be updated in order to mitigate the noise and interference.
In the open literature, there are four main write-voltage design techniques adopted in the flash memory.
The first three techniques are the fixed write-voltage design scheme \cite{1888888},  MCC write-voltage design scheme \cite{2000000} and minimum-RBER write-voltage design scheme \cite{9169701}.
Nevertheless, the above three write-voltage design schemes do not consider the asymmetric property of RBERs \cite{6804933, 9000906}.
The last one is the MRD write-voltage design scheme \cite{8075860}, but it does not consider the influence of LDPC code on flash memory system.
To address these issue, a dynamic write-voltage design scheme, which takes the unbalanced RBERs between MSB and LSB pages as well as the LDPC codes into consideration, is proposed in this paper.

According to \cite{5629456}, there is a nonlinear relation between BER and RBER, which means that this relationship can be expressed by a function.
However, when using the BP decoding, it is impossible to get the BER from the RBER through theoretical derivation.
Because Maximum-likelihood (ML) decoding is an optimal decoding scheme \cite{8735878}, the BER under ML decoding can be treated as the lower bound of the BER under any decoding.
Thus, we use the following two formulas based on the ML decoding to estimate the BERs of LSB and MSB pages [20, Eq.~(15) and Eq.~(16)], i.e.,
\begin{eqnarray}\label{eq:BER_formula_LSB}
P_{\rm LSB}^{\rm ML} \approx& [(\alpha_{\rm 01}+\alpha_{\rm 23})\omega]^{\lceil {\frac {d_2^{\rm min}} 2} \rceil} \nonumber \\
\times&\sum\limits_{d_2 \in \{ d_2^{\rm min}, 2\lfloor {\frac {d_2^{\rm min}+1} 2} \rfloor \}}A_2(d_2) 2^{-\lceil {\frac {3 d_2} 2} \rceil} {\binom {d_2} {\lceil {d_2}/2 \rceil}} , \nonumber \\
\end{eqnarray}
\begin{eqnarray}\label{eq:BER_formula_MSB}
P_{\rm MSB}^{\rm ML} \approx& (\alpha_{\rm 12}\omega)^{\lceil {\frac {d_1^{\rm min}} 2} \rceil} \nonumber \\
\times&\sum\limits_{d_1 \in \{ d_1^{\rm min}, 2\lfloor {\frac {d_1^{\rm min}+1} 2} \rfloor \}}A_1(d_1) 4^{-{d_1}} {\binom {d_1} {\lceil {d_1}/2 \rceil}} , \nonumber \\
\end{eqnarray}
where $\alpha_{gh}\omega$ is the transition probability from voltage state $g$ to $h$, and $g, h \in \{ 0, 1, 2, 3 \}$.
There is a mapping rule between $g/h$ and the voltage state, i.e.,
$\{ 0\rightarrow 11, 1\rightarrow 10, 2\rightarrow 00, 3\rightarrow 01 \}$.
For the LDPC code employed by page $z$,
$d_z$ is the code distance and $A_z(d_z)$ is the Hamming distance spectrum, where $z=1$ and $2$ denote the MSB page and LSB page, respectively.

Because MSB and LSB pages use the same LDPC code, and the corresponding minimum codeword distances $d_z^{\rm min}$ on both pages are identical, A simplified function (i.e., cost function \eqref{eq:write-cost}) can be formulated by removing the terms involving $d_z$ (i.e., $A_z(d_z)$ and ${\binom {d_z} {\lceil {d_z}/2 \rceil}}$) from Eq.~\eqref{eq:BER_formula_LSB} and Eq.~\eqref{eq:BER_formula_MSB}.
The above simplification can significantly reduce the computational overhead.

In this cost function, we divide the RBER into two parts,
the RBERs of MSB and LSB pages. Finally, the write voltage is determined by the cost-function output.
In this work, by considering the asymmetric property of the MSB-page and LSB-page error rates,
as well as the minimum Hamming distance of LDPC code,
we first define the cost function $\mathit{C}_{\rm write}$ for an MLC flash memory channel as follows:
\begin{eqnarray}\label{eq:write-cost}
\mathit{C}_{\rm write} = 2^{(-{\frac 3 2}d^{\rm min})} \omega_{\rm lsb}+4^{(-d^{\rm min})}\omega_{\rm msb},
\end{eqnarray}
where $d^{\rm min}$ is the minimum Hamming distance of the LDPC code that can be estimated by \cite{9567703},
$\omega_{\rm lsb}$ and $\omega_{\rm msb}$ are the RBERs of LSB and MSB pages, respectively.
Assume that the MLC flash memory exploits three read voltages {\color{blue}$\{t_1^*, t_2^*, t_3^*\}$}
to distinguish four threshold-voltage states, and the input probabilities of four data symbols are equal, i.e., $p_i=1/4$.
Therefore, we can compute the RBERs of MSB and LSB pages as \cite{9169701}
\begin{eqnarray}\label{eq:MSB-Pe}
\omega_{\rm msb} = {\frac 1 4}\left[{p_{s_{10}}(v>{t_2^*})}+{p_{s_{00}}(v<{t_2^*})}\right],
\end{eqnarray}
\begin{eqnarray}\label{eq:LSB-Pe}
\omega_{\rm lsb} = {\frac 1 4}\left[p_{s_{11}}(v>t_1^*)+p_{s_{10}}(v<t_1^*)\right] \nonumber \\
+{\frac 1 4}\left[p_{s_{00}}(v>t_3^*)+p_{s_{01}}(v<t_3^*)\right],
\end{eqnarray}
where $t_1^*$, $t_2^*$ and $t_3^*$ are read-voltage levels at the intersections of adjacent voltage states.
For the sake of obtaining $t_1^*, t_2^*$ and $t_3^*$, we need to solve three equations:
\begin{eqnarray}\label{eq:three_equations}
p_{s_{11}}(v=t_1^*) = p_{s_{10}}(v=t_1^*) \nonumber \\
p_{s_{10}}(v=t_2^*) = p_{s_{00}}(v=t_2^*) \nonumber \\
p_{s_{00}}(v=t_3^*) = p_{s_{01}}(v=t_3^*) . \nonumber
\end{eqnarray}We get $t_1^* \in (V_{\rm min}, V_1)$, $t_2^* \in (V_1, V_2)$ and $t_3^* \in (V_2, V_{\rm max})$.
Since Eq.~\eqref{eq:write-cost} represents the cost of BER,
we need to find $V_1^*$ and $V_2^*$ that minimize the output of Eq.~\eqref{eq:write-cost}.
As a result, the optimal $V_1^*$ and $V_2^*$ can be yielded by tackling with the following optimization problem
\begin{align}\label{eq:write-min}
(V_1^*, V_2^*) = \underset {(V_1, V_2)}{\arg\min} \mathit{C}_{\rm write}.
\end{align}

Based on the above discussion, we first search for $V_1^*$ by fixing $V_2^*$ to minimize the cost function.
Since $V_1 \in (V_{\rm min}, V_2)$, we can uniformly divide this region into $M-1$ intervals, and thus obtain the boundaries ${\boldsymbol V}_1 = \{V_1^1, V_1^2, \ldots, V_1^M\}$ of those intervals.
To get the optimal $V_1^*$, we need to obtain all the values of cost function corresponding to $M$ write voltages.
Through such a method, the optimal $V_1^*$ corresponding to the minimum value of $\mathit{C}_{\rm write}$ can be obtained by employing the bisection search method \cite{2222222}.
Likewise, the optimal value $V_2^*$ can be also obtained by fixing $V_1^*$, where $V_2 \in (V_1, V_{\rm max})$.
These two search operations are repeated in a sequential order until $V_1^*$ and $V_2^*$ no longer change or the maximum number $q$ of iteration is reached.
The details of the proposed dynamic write-voltage design scheme are summarized in Algorithm 1.

\begin{algorithm}[t]
\label{write-voltage algorithm split}
{\small \caption{Proposed write-voltage design algorithm}}
\begin{algorithmic}[1]
\STATE {\bf Initialization}: $V_2^* = 3.3$. \\
\FOR{$q = 1$ {\rm to} $50$}
    \FOR{$V_1 \in (V_{\rm min}, V_2)$}
        \STATE Set $V_2 = V_2^*$; \\
        \STATE Uniformly divide ($V_{\rm min}, V_2$) into ${M-1}$ intervals; \\
        \STATE Calculate the RBERs of MSB and LSB pages by applying Eq.~\eqref{eq:MSB-Pe} and Eq.~\eqref{eq:LSB-Pe}; \\
        \STATE Calculate the values of cost function corresponding to $M$ write voltages by using Eq.~\eqref{eq:write-cost};
    \ENDFOR
    \STATE Find {$V_1 \in (V_{\rm min}, V_2)$} that minimizes $\mathit{C}_{\rm write}$; \\
    \STATE Set $V_1^* = V_1$; \\
    \FOR{$V_2 \in (V_1, V_{\rm max})$}
        \STATE Set $V_1 = V_1^*$; \\
        \STATE Uniformly divide ($V_1, V_{\rm max}$) into $M-1$ intervals; \\
        \STATE Calculate the RBERs of MSB and LSB pages by applying Eq.~\eqref{eq:MSB-Pe} and Eq.~\eqref{eq:LSB-Pe}; \\
        \STATE Derives the values of cost function corresponding to $M$ write voltages by using Eq.~\eqref{eq:write-cost};
    \ENDFOR
    \STATE Find {$V_2 \in (V_1, V_{\rm max})$} that minimizes $\mathit{C}_{\rm write}$; \\
    \STATE Set $V_2^* = V_2$. \\
    \IF {$V_1^*$ and $V_2^*$ no longer change}
    \STATE {\bf break;}
    \ENDIF
\ENDFOR
\end{algorithmic}
\end{algorithm}

\section{Proposed Cost-Function-Based Read-Voltage Optimization Scheme}\label{read-thresholds-optimization}

\subsection{Preliminaries}
In \cite{9169701}, an entropy-based quantization scheme was proposed to design the read voltage by varying the entropy $\theta$.
Nevertheless, this method obtains the optimal entropy $\theta^* = 0.35$ by fixing the retention time as $0$.
Once the retention time $T$ varies, the entropy-based quantization scheme needs to use a large number of decoding operations for acquiring the optimal $\theta^*$, inevitably causing high computational complexity.
To address this issue, a novel read-voltage optimization scheme is proposed in this section, which takes the inaccurate LLRs and the unbalanced RBERs of the LDPC-coded flash memory system into consideration.
Our objective is to obtain the optimal read voltage while maintaining low computational complexity.

To begin with, we define the entropy function $H(v)$ as \cite{9169701}
\begin{align}\label{eq:Entropy-Function}
H(v) = \sum_{i} \left[ \frac {p_{s_i}(v)} {\sum\limits_i p_{s_i}(v)} \log_2 \left(\frac {\sum\limits_{i}p_{s_i}(v)} {p_{s_i}(v)} \right) \right],
\end{align}
where $s_i \in \{s_{11}, s_{10}, s_{00}, s_{01}\}$ and $p_{s_i}$ is the PDF of the threshold-voltage state $s_i$.

The main error area usually occurs in the overlapping areas between adjacent threshold-voltage states.
Thus, we can optimize the read voltage to determine the main error area by resolving the following function \cite{9169701}
\begin{align}\label{eq:theta}
H(R_n) = \theta,~~~~\theta \in [0, 1].
\end{align}
Thus, we can obtain six read voltages, i.e., $R_1, R_2, \ldots, R_6$, by changing the value of $\theta$.
The optimization of read voltage can be transformed to the optimization of $\theta$.

Moreover, as shown in Fig.~\ref{Fig1}, we can calculate the LLR information for each threshold-voltage region, as \cite{9169701, 2555555}
\begin{align}\label{eq:LLR-MSB}
L_{\rm msb} &=  \log{\frac  {\int_{R_{n-1}}^{R_n} \{p_{s_{00}}(v)+p_{s_{01}}(v)\}dv} {\int_{R_{n-1}}^{R_n} \{p_{s_{10}}(v)+p_{s_{11}}(v)\}dv}},  \\
\label{eq:LLR-LSB}
L_{\rm lsb} &=  \log{\frac  {\int_{R_{n-1}}^{R_n} \{p_{s_{00}}(v)+p_{s_{10}}(v)\}dv} {\int_{R_{n-1}}^{R_n} \{p_{s_{01}}(v)+p_{s_{11}}(v)\}dv}},
\end{align}
where $R_n (n=1, 2,\ldots,6)$ is the $n$-th read voltage, $R_{n-1}< v <R_n$.

\subsection{Proposed Optimization Scheme}

Considering the effect of the inaccurate LLR information and the unbalanced RBERs, we propose to optimize the entropy $\theta$ by minimizing a new cost function.

When three optimal read voltages $t_1^*, t_2^*$ and $t_3^*$ are used for soft-decision BP decoding,
two LLR sequences corresponding to LDPC codewords can be obtained,
which belong to LSB and MSB pages, respectively.
Then,
the BER of LDPC codes under ML decoding can be estimated \cite{6403864},
which is similar to Eq.~\eqref{eq:BER_formula_LSB} and Eq.~\eqref{eq:BER_formula_MSB}.

When the entropy-based quantization scheme is used,
the read voltages will be set near the intersections (i.e., $t_1^*$, $t_2^*$ and $t_3^*$) of adjacent voltage states.
If the threshold voltage is in a region including an intersection, its corresponding LLR value is close to 0.
Exploiting ML decoding, the LLR value 0 can be decoded as either $+\zeta$ or $-\zeta$ with a high probability, where $\zeta$ is a constant. As can be observed, 0 has the same distance to $+\zeta$ and $-\zeta$, which does not affect the result of ML decoding.
Thus, we can use the BER of LDPC codes using three-optimal read voltages under ML decoding to find the optimal entropy of the entropy-based quantization scheme.

The read voltages are directly related to BP decoding, because the LLR values of all threshold-voltage regions can be calculated after determining the read voltages.
In this sense, the change of LLR value
when error occurs can better reflect the noise level as $\omega_{\rm lsb}$ and $\omega_{\rm msb}$ in the Eq.~\eqref{eq:write-cost}.

The proposed read-voltage scheme is different from that in \cite{8735878} because the latter assumes that the RBERs of bit 1 and bit 0 are equal, while the former assumes that they are unequal. In practical flash memory, the RBERs of bit 1 and bit 0 are always unequal. One can measure this inequality level by calculating the expected LLR of error bits.

Therefore, using the equations \eqref{eq:MSB-Pe}, \eqref{eq:LSB-Pe}, \eqref{eq:LLR-MSB}, and \eqref{eq:LLR-LSB},
we can compute the noise level $\alpha_{\rm lsb}^{P_e}$ and
the expected LLR $\alpha_{\rm lsb}^{\rm llr}$ corresponding to the error bits in the LSB page under BP decoding, given by
\begin{eqnarray}\label{eq:error_related_value2}
\alpha_{\rm lsb}^{P_e} &= {\sum\limits_{i=1}^7}|L_{\rm lsb}^i|\cdot P_{\rm lsb}^i,
\end{eqnarray}
\begin{eqnarray}\label{eq:error_related_value1}
\alpha_{\rm lsb}^{\rm llr} &= {\sum\limits_{i=1}^7}L_{\rm lsb}^i\cdot P_{\rm lsb}^i,
\end{eqnarray}
where $L_{\rm lsb}^i$ represents the corresponding LLR value of LSB on the interval [$R_{i-1}$, $R_i$], $(i=1, 2,\ldots,7)$,
as shown in Fig.~\ref{Fig1}.
In general, $R_0$ and $R_7$ are set as $-\infty$ and $+\infty$.
$P_{\rm lsb}^i$ is the RBER of LSB page in the interval [$R_{i-1}$, $R_i$].
In addition, $\alpha_{\rm msb}^{P_e}$ and $\alpha_{\rm msb}^{\rm llr}$ can be calculated in similar way.

Similar to the cost function \eqref{eq:write-cost}, the cost functions \eqref{eq:read_cost_1} and \eqref{eq:read_cost_2} can be constructed according to the input variables $\alpha_{\rm lsb}^{P_e}$ and $\alpha_{\rm lsb}^{\rm llr}$, respectively, i.e.,
\begin{eqnarray}\label{eq:read_cost_1}
\mathit{C}_{\rm read}^{P_e} = 2^{(-{\frac 3 2} d^{\rm min})}\alpha_{\rm lsb}^{P_e}+4^{(-d^{\rm min})}\alpha_{\rm msb}^{P_e},
\end{eqnarray}
\begin{eqnarray}\label{eq:read_cost_2}
\mathit{C}_{\rm read}^{\rm llr} = 2^{(-{\frac 3 2} d^{\rm min})} \alpha_{\rm lsb}^{\rm llr}+4^{(-d^{\rm min})}\alpha_{\rm msb}^{\rm llr},
\end{eqnarray}where $d^{\rm min}$ is the minimum Hamming distance of the LDPC code that can be estimated by \cite{9567703},
$\alpha_{\rm lsb}^{P_e}$, $\alpha_{\rm msb}^{P_e}$, $\alpha_{\rm lsb}^{\rm llr}$ and $\alpha_{\rm msb}^{\rm llr}$ are given in the Eq.~\eqref{eq:error_related_value2} and Eq.~\eqref{eq:error_related_value1}.

In order to comprehensively consider the impact of the noise level and expected LLR of error bits on the BER,
we can obtain the overall cost function, as
\begin{eqnarray}\label{eq:read_cost_final}
\mathit{C}_{\rm read}^{\rm oa} = \mathit{c}_1\cdot\mathit{C}_{\rm read}^{P_e} + \mathit{c}_2\cdot\mathit{C}_{\rm read}^{\rm llr}.
\end{eqnarray}

In fact, the effect of the noise level and expected LLR on the BER performance of an LDPC-coded flash memory is uncertain.
For this reason, two weighted factors $\mathit{c}_1$ and $\mathit{c}_2$ need to be included in the Eq.~\eqref{eq:read_cost_final}
in order to indicate their influence on the BER performance.
To estimate the weighted factors $\mathit{c}_1$ and $\mathit{c}_2$,
we first need to obtain a BER sequence versus the entropy $\theta$ under the BP decoding.
Subsequently, we use the linear regression technology to couple the cost function and BER \cite{2666666}.
Finally, the two weighted factors are yielded.

With the help of the LLR-aware cost function \eqref{eq:read_cost_final},
the optimal entropy $\theta^*$ can be derived by solving the following optimization problem
\begin{align}\label{eq:read-theta-optimal}
\theta^* = \underset {\theta}{\arg\min} \mathit{C}_{\rm read}^{\rm oa}.
\end{align}

\begin{figure}[t]
\centering
\subfloat[cost function]{
\includegraphics[width=0.45\textwidth]{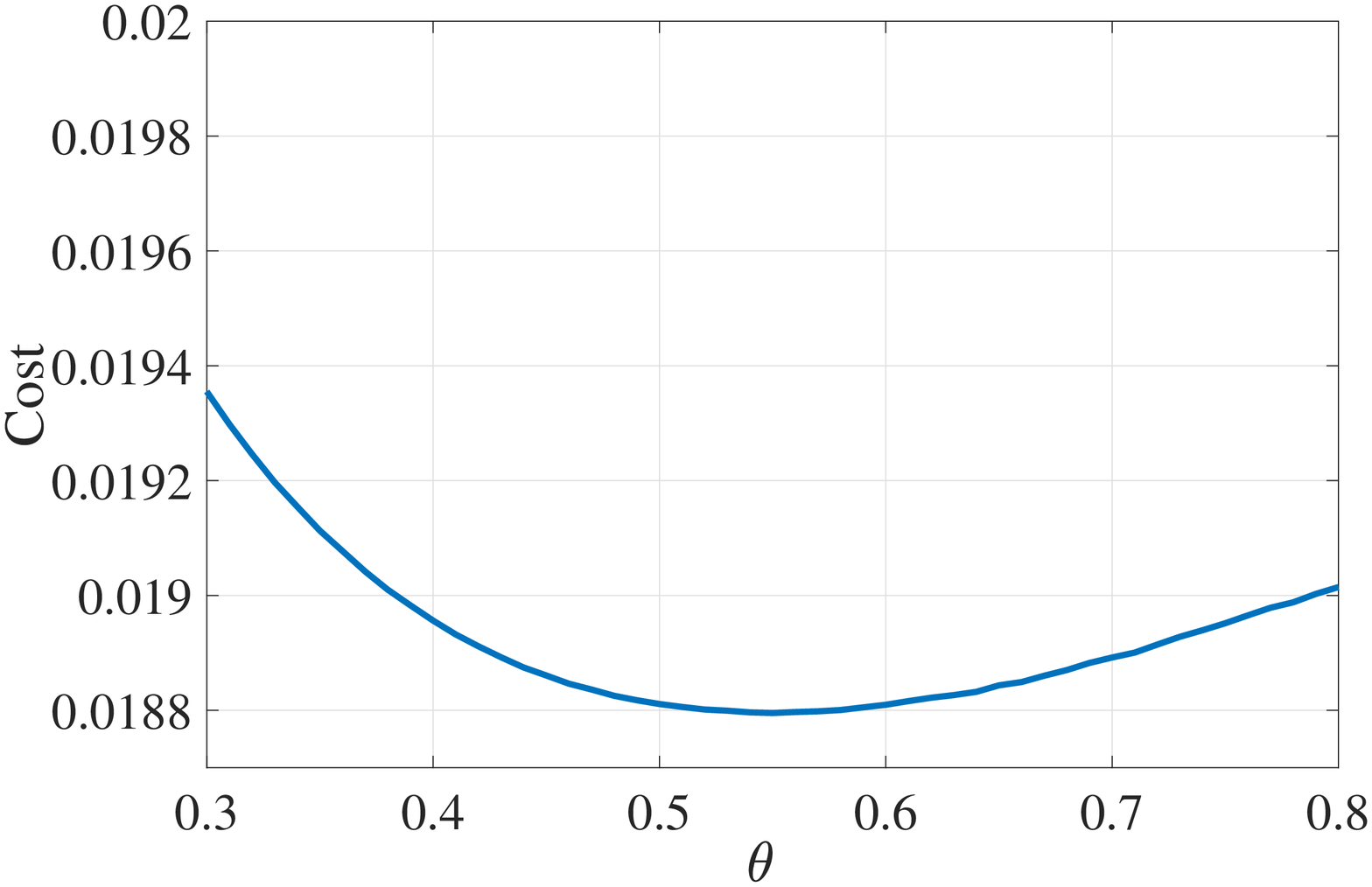}
\label{fig:Fig2_a}
}

\subfloat[BER performance]{
\includegraphics[width=0.45\textwidth]{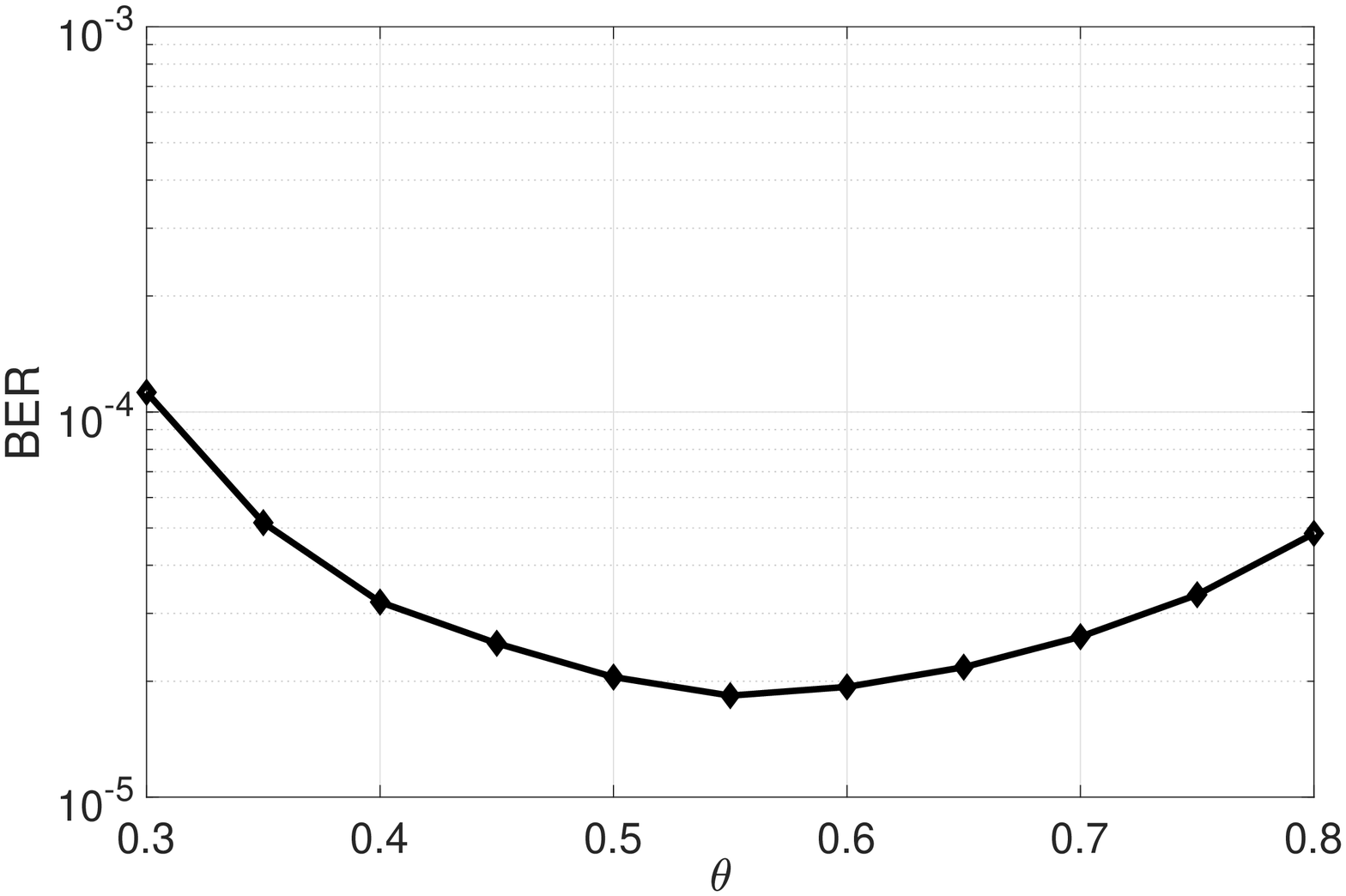}
\label{fig:Fig2_b}
}
\caption{The corresponding cost function value (a) and BER performance (b) of the proposed read-voltage optimization scheme versus the entropy $\theta$ over an MLC flash memory channel, where the channel coding is implemented by a rate-$0.89$ protograph LDPC code \cite{6517051},
the PE cycles is set to $6000$ and the retention time $T$ is set to $15000$.}
\label{fig:Fig2}
\end{figure}

Fig.~\ref{fig:Fig2_a} depicts the cost function by varying the value of $\theta$.
As shown, the cost function is concave function, the gradient-descent (GD) algorithm \cite{2333333} can be used to search for the optimal $\theta^*$ that yields the minimum value of $\mathit{C}_{\rm read}^{\rm oa}$.
Then, we can obtain the optimal $\{R_1^*,R_2^*,\ldots,R_6^*\}$ corresponding to the optimal $\theta^*$.
To guarantee the desirable BER performance of flash memory,
the cost function is required to be re-calculated so as to get the new optimal read voltages when the PE cycles or retention time varies.

\begin{table*}[tb]\scriptsize
\centering
\caption{Number of operations required for calculating a single entropy $\theta$.}
\label{table1}
\begin{tabular}{|c|c|c|c|c|}
\hline
                                         & Number of addition & Number of multiplications & Number of special operations  \\ \hline
Cost function-based quantization scheme  & $4(Q+1)+3$         & $4(Q+1)+6$                & 0                             \\ \hline
Entropy-based quantization scheme  \cite{9169701}  & $2(2{d_v}+1)IB$    & $4{d_c}I{\eta}B$          & $2{d_c}I{\eta}B$ $\tanh(x)$, $2{d_c}I{\eta}B$ $\tanh^{-1}(x)$            \\ \hline
\end{tabular}
\end{table*}

\subsection{Verification of Proposed Optimization Scheme }

Fig.~\ref{fig:Fig2} shows the BER performance and the corresponding cost function value of the proposed read-voltage optimization scheme versus the entropy $\theta$, where the PE cycles is set to $6000$ and the retention time $T$ is set to $15000$.
It can be seen that the minimum BER performance
of the proposed read-voltage optimization scheme also can be obtained at the optimal $\theta^*=0.55$ that yields the minimum value of the cost function.
We have also performed verification with the same parameter setting as in \cite{9169701}, and have obtained the same optimal entropy value as $\theta^*=0.35$.
The above phenomenon verifies that the proposed read-voltage optimization scheme can derive the optimal entropy $\theta^*$ achieving the minimum BER performance.

\subsection{Complexity Analysis}
The proposed read-voltage optimization scheme is based on an LLR-aware cost function, while the entropy-based read-voltage optimization scheme in \cite{9169701} is based on the output of BP decoding.
Here, we briefly compare their computational overheads so as to further illustrate the superiority of our design.
When comparing the computational overhead, we do not consider the estimation of threshold-voltage distribution and associated LLR values, because these two steps are involved in both schemes.
Now, we set $Q$ to be the number of read voltages. For the proposed scheme, the computational overhead mainly comes from the measurement of the variables $\alpha_{\rm lsb}^{P_e}$, $\alpha_{\rm lsb}^{\rm llr}$, $\alpha_{\rm msb}^{P_e}$, $\alpha_{\rm msb}^{\rm llr}$ and $\mathit{C}_{\rm read}^{\rm oa}$.
As such, the computational overhead of the proposed cost function-based quantization scheme is ${\cal O}(8(Q+1))$ for obtaining a single entropy $\theta$.

On the other hand, the computational overhead of the entropy-based quantization scheme mainly comes from the LLR update of variable nodes and check nodes within an LDPC code \cite{1495850,8338131,9586047}.
We set $d_v$ to be the average degree of variable nodes of the protograph LDPC code, $d_c$ to be the average degree of check nodes, $\eta$ to be the number of check nodes, $N$ to be the length of the codeword sequence, $I$ to be the average decoding-iteration number for each codeword, and $B$ is the number of codewords generated in BP decoding simulation for calculating a single entropy $\theta$ (e.g., when the frame error rate is equal to $10^{-6}$, the number of codewords should be greater than $10^{8}$ to obtain a sufficiently reliable simulation result).
Therefore, the computational overhead of the entropy-based quantization scheme is ${\cal O}(2(2{d_v}+1)IB+4{d_c}I{\eta}B+4{d_c}I{\eta}B)$ $\approx$ ${\cal O}(8{d_c}I{\eta}B)$ because $\eta \gg 2$.

Table \ref{table1} summarizes the computational overheads of the proposed cost function-based quantization scheme and the entropy-based quantization scheme required for calculating a single entropy $\theta$. As observed, the proposed read-voltage optimization scheme benefits from much lower computational complexity with respect to entropy-based quantization scheme.
Besides, the proposed read-voltage optimization scheme and the entropy-based quantization scheme \cite{9169701} should take the same storage space to store the entropy values and their corresponding cost function values and BERs, respectively.

In application, the optimal read voltages at a certain PE cycles and retention time can be calculated off-line,
and then stored into a look-up table in the flash memory.

\section{Simulation Results}\label{Simulation Results}

In this section, we present various simulation results of the proposed dynamic write-voltage design scheme and the read-voltage optimization scheme to verify their superiority over the flash memory channel.
The simulations are carried out by using MATLAB. We utilize the rate-$0.89$ protograph LDPC code \cite{6517051, 9600574} in the simulations, which is constructed by employing a modified progressive-edge-growth (PEG) algorithm \cite{2444444}.
For decoding of LDPC codes, we use the BP algorithm and assume the maximum number of iterations as $50$.

\subsection{BER Performance of Dynamic Write-Voltage Design Scheme}
\begin{figure}[h]
\centering
\includegraphics[width=0.5\textwidth]{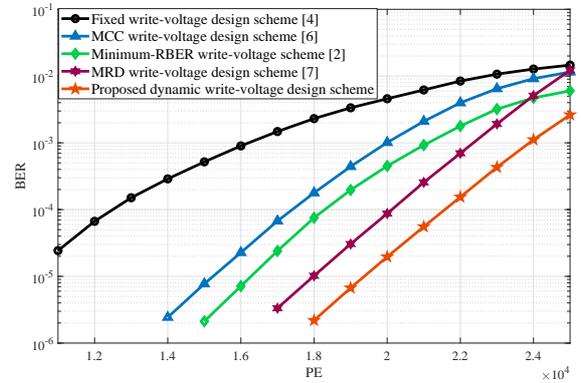}
\caption{BER performance of the fixed write-voltage design scheme, MCC write-voltage design scheme, minimum-RBER write-voltage design scheme, MRD write-voltage design scheme and the
proposed dynamic write-voltage design scheme versus the PE cycles.}
\label{fig:Fig3}
\end{figure}

\begin{figure}[h]
\centering
\includegraphics[width=0.5\textwidth]{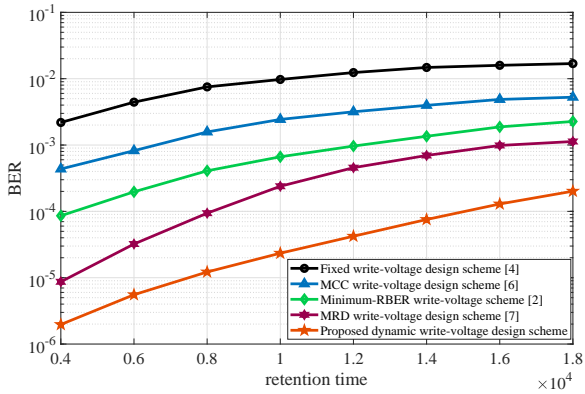}
\caption{BER performance of the fixed write-voltage design scheme, MCC write-voltage design scheme, minimum-RBER write-voltage design scheme, MRD write-voltage design scheme and the
proposed dynamic write-voltage design scheme versus the retention time.}
\label{fig:Fig3_retention}
\end{figure}

Fig.~\ref{fig:Fig3} and {\color{blue}Fig.~\ref{fig:Fig3_retention}} show the BER performance of the fixed write-voltage design scheme \cite{1888888}, MCC write-voltage design scheme \cite{2000000}, minimum-RBER write-voltage design scheme \cite{9169701}, MRD write-voltage design scheme \cite{8075860} and the
proposed dynamic write-voltage design scheme versus the PE cycles and retention time over an MLC flash memory channel, respectively.
It can be seen that the performance of the proposed dynamic write-voltage design scheme is obviously superior to that of the other four write-voltage design schemes.
It is because that we substantially consider the unbalanced RBERs between MSB and LSB pages.
In particular, at ${\rm PE} =18000$, the proposed dynamic write-voltage design scheme achieves a BER of $2.2 \times 10^{-6}$, while the MRD write-voltage design scheme \cite{8075860}, minimum-RBER write-voltage design scheme \cite{9169701}, MCC write-voltage design scheme \cite{2000000} and the fixed write-voltage design scheme \cite{1888888} only accomplish the BERs of $1.0 \times 10^{-5}$, $7.5 \times 10^{-5}$, $1.8 \times 10^{-4}$ and $2.3 \times 10^{-3}$, respectively.

\subsection{ BER Performance of Read-Voltage Optimization Scheme}

\begin{figure}[t]
\center
\includegraphics[width=0.5\textwidth]{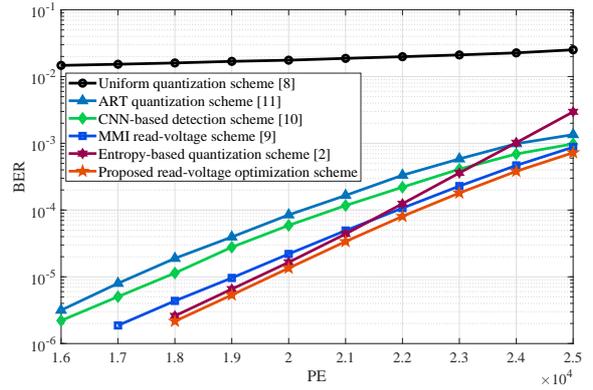}
\caption{BER performance of the uniform quantization scheme, ART quantization scheme, CNN-based detection scheme, MMI read-voltage scheme, entropy-based quantization scheme $(\theta = 0.35)$ and proposed read-voltage optimization scheme versus the PE cycles.}
\label{fig:Fig4}
\end{figure}

\begin{figure}[t]
\center
\includegraphics[width=0.5\textwidth]{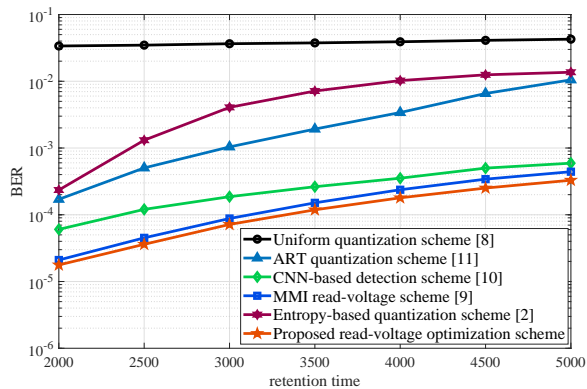}
\caption{BER performance of the uniform quantization scheme, ART quantization scheme, CNN-based detection scheme, MMI read-voltage scheme, entropy-based quantization scheme $(\theta = 0.35)$ and proposed read-voltage optimization scheme versus the retention time.}
\label{fig:Fig5}
\end{figure}

Fig.~\ref{fig:Fig4} and Fig.~\ref{fig:Fig5} show the BER performance of the uniform quantization scheme \cite{8314735}, ART quantization scheme \cite{7152879}, CNN-based detection scheme \cite{2555555}, MMI read-voltage quantization scheme \cite{8708250}, entropy-based quantization scheme \cite{9169701} and proposed read-voltage optimization scheme versus the PE cycles and retention time over an MLC flash memory channel, respectively.
As shown, the proposed read-voltage optimization scheme not only slightly outperforms the MMI quantization scheme, but also is
significantly superior to the CNN-based detection scheme, the ART quantization scheme, the entropy-based quantization scheme and the uniform quantization scheme.
For example, at ${\rm retention~time} = 2500$, the proposed read-voltage optimization scheme obtains a BER of $3.6 \times 10^{-5}$, while the MMI read-voltage quantization scheme, the CNN-based detection scheme, the ART quantization scheme, the entropy-based quantization scheme, and the uniform quantization scheme obtain the BERs of $4.5 \times 10^{-5}$, $1.2 \times 10^{-4}$, $5.0 \times 10^{-4}$, $1.3 \times 10^{-3}$ and $3.5 \times 10^{-2}$, respectively.

\section{Conclusions}\label{Conclusions}
To improve the reliability of the LDPC-coded MLC flash memory, we have presented a novel dynamic write-voltage design scheme to update the write voltage by considering the asymmetric property of the MSB-page and LSB-page error rates.
Besides, we have conceived an LLR-aware cost function to optimize the read voltages, which can further improve the BER performance of flash memory with relatively low computational complexity.
Simulation results have shown that our proposed dynamic write-voltage design and read-voltage optimization scheme outperforms
the state-of-the-art schemes over MLC flash memory channel.
Besides, the proposed schemes can be extended to 3D NAND flash memory after proper modifications.

\bibliographystyle{IEEEtran}
\bibliography{myref}

\begin{thebibliography}{10}
\providecommand{\url}[1]{#1}
\csname url@samestyle\endcsname
\providecommand{\newblock}{\relax}
\providecommand{\bibinfo}[2]{#2}
\providecommand{\BIBentrySTDinterwordspacing}{\spaceskip=0pt\relax}
\providecommand{\BIBentryALTinterwordstretchfactor}{4}
\providecommand{\BIBentryALTinterwordspacing}{\spaceskip=\fontdimen2\font plus
\BIBentryALTinterwordstretchfactor\fontdimen3\font minus
  \fontdimen4\font\relax}
\providecommand{\BIBforeignlanguage}[2]{{%
\expandafter\ifx\csname l@#1\endcsname\relax
\typeout{** WARNING: IEEEtran.bst: No hyphenation pattern has been}%
\typeout{** loaded for the language `#1'. Using the pattern for}%
\typeout{** the default language instead.}%
\else
\language=\csname l@#1\endcsname
\fi
#2}}
\providecommand{\BIBdecl}{\relax}
\BIBdecl

\bibitem{9264190}
D.~{Wu}, H.~{You}, X.~{Wang}, S.~{Zhong}, and Q.~{Sun}, ``Experimental
  investigation of threshold voltage temperature effect during
  cross-temperature write–read operations in {3-D} {NAND} flash,'' \emph{IEEE
  J. Electron Devices Soc.}, vol.~9, pp. 22--26, 2021.

\bibitem{9169701}
C.~A. {Aslam}, Y.~L. {Guan}, and K.~{Cai}, ``Read and write voltage signal
  optimization for multi-level-cell ({MLC}) {NAND} flash memory,'' \emph{IEEE
  Trans. Commun.}, vol.~64, no.~4, pp. 1613--1623, Apr. 2016.

\bibitem{8351503}
A.~K. {Subbiah} and T.~{Ogunfunmi}, ``Area-effcient re-encoding scheme for
  {NAND} flash memory with multimode {BCH} error correction,'' in \emph{Proc.
  IEEE. Int. Symp. Circuits Syst. (ISCAS)}, May 2018, pp. 1--5.

\bibitem{1888888}
C.~A. {Aslam}, Y.~L. {Guan}, and K.~{Cai}, ``Dynamic write-level and read-level
  signal design for {MLC} {NAND} flash memory,'' in \emph{Proc. 9th Int. Symp.
  Commun. Syst. Netw. Digit. Sign(CSNDSP)}, Jul. 2014, pp. 336--341.

\bibitem{1999999}
Y.~{Kim}, J.~{Kim}, J.~J. {Kong}, B.~{K Vijaya Kumar}, and X.~{Li}, ``Verify
  level control criteria for multi-level cell flash memories and their
  applications,'' \emph{EURASIP J. Adv. Signal Process}, vol. 2012, no.~1, pp.
  1--13, 2012.

\bibitem{2000000}
C.~{Duangthong}, W.~{Phakphisut}, and P.~{Supnithi}, ``Capacity enhancement of
  asymmetric multi-level cell ({MLC}) {NAND} flash memory using write voltage
  optimization,'' in \emph{Proc. ITC-CSCC}, jun. 2019, pp. 1--4.

\bibitem{8075860}
C.~{D}uangthong, W.~{P}hakphisut, and P.~{S}upnithi, ``Search algorithm of
  write voltage optimization in {NAND} flash memory,'' in \emph{Proc. Int.
  Electr. Eng. Congr. (iEEECON)}, Mar. 2017, pp. 1--4.

\bibitem{8314735}
G.~{Dong}, N.~{Xie}, and T.~{Zhang}, ``On the use of soft-decision
  error-correction codes in {NAND} flash memory,'' \emph{IEEE Trans. Circuits
  Syst. I: Regul. Papers}, vol.~58, no.~2, pp. 429--439, Feb. 2010.

\bibitem{8708250}
J.~{Wang}, K.~{Vakilinia}, T.-Y. {Chen}, T.~{Courtade}, G.~{Dong}, T.~{Zhang},
  H.~{Shankar}, and R.~{Wesel}, ``Enhanced precision through multiple reads for
  {LDPC} decoding in flash memories,'' \emph{IEEE J. Sel. Areas Commun.},
  vol.~32, no.~5, pp. 880--891, May 2014.

\bibitem{2555555}
Z.~F. {Shi}, Y.~{Fang}, Y.~C. {Bu}, and G.~J. {Han}, ``Convolutional neural
  network ({CNN})-based detection for multi-level-cell {NAND} flash memory,''
  \emph{IEEE Commun. Lett.}, vol.~25, no.~12, pp. 3883--3887, Dec. 2021.

\bibitem{7152879}
B.~{Peleato}, R.~{Agarwal}, J.~M. {Cioffi}, M.~{Qin}, and P.~H. {Siegel},
  ``Adaptive read thresholds for {NAND} flash,'' \emph{IEEE Trans. Commun.},
  vol.~63, no.~9, pp. 3069--3081, Sept. 2015.

\bibitem{6484076}
H.~{Wang}, N.~{Wong}, T.-Y. {Chen}, and R.~D. {Wesel}, ``Using dynamic
  allocation of write voltage to extend flash memory lifetime,'' \emph{IEEE
  Trans. Commun.}, vol.~64, no.~11, pp. 4474--4486, Nov. 2016.

\bibitem{8819688}
K.-D. {Suh}, B.-H. {Suh}, Y.-H. {Lim}, J.-K. {Kim}, Y.-J. {Choi}, Y.-N. {Koh},
  S.-S. {Lee}, S.-C. {Kwon}, B.-S. {Choi}, J.-S. {Yum} \emph{et~al.}, ``A 3.3
  {V} 32 {Mb} {NAND} flash memory with incremental step pulse programming
  scheme,'' \emph{IEEE J. Solid-State Circuits}, vol.~30, no.~11, pp.
  1149--1156, Nov. 1995.

\bibitem{9416922}
Z.~{Fang}, Z.~{Ma}, X.~{Tang}, Y.~{Xiao}, and Y.~{Tang}, ``Program error
  mitigation in {MLC} {NAND} flash memory with soft decision decoders,''
  \emph{China Commun.}, vol.~18, no.~4, pp. 76--87, Apr. 2021.

\bibitem{2111111}
G.~{Dong}, N.~{Xie}, and T.~{Zhang}, ``Enabling {NAND} flash memory use
  soft-decision error correction codes at minimal read latency overhead,''
  \emph{IEEE Trans. Circuits Syst. I: Regul. Papers}, vol.~60, no.~9, pp.
  2412--2421, Sept. 2013.

\bibitem{5460923}
G.~{Dong}, S.~{Li}, and T.~{Zhang}, ``Using data postcompensation and
  predistortion to tolerate cell-to-cell interference in {MLC} {NAND} flash
  memory,'' \emph{IEEE Trans. Circuits Syst. I: Reg. Papers}, vol.~57, no.~10,
  pp. 2718--2728, Oct. 2010.

\bibitem{6804933}
P.~{Chen}, K.~{Cai}, and S.~{Zheng}, ``Rate-adaptive protograph {LDPC} codes
  for multi-level-cell {NAND} flash memory,'' \emph{IEEE Commun. Lett.},
  vol.~22, no.~6, pp. 1112--1115, Jun. 2018.

\bibitem{9000906}
F.~{Wu}, M.~{Zhang}, Y.~{Du}, W.~{Liu}, Z.~{Lu}, J.~{Wan}, Z.~{Tan}, and
  C.~{Xie}, ``Using error modes aware {LDPC} to improve decoding performance of
  {3-D} {TLC} {NAND} flash,'' \emph{IEEE Trans. Comput.-Aided Des. Integr.
  Circuits Syst.}, vol.~39, no.~4, pp. 909--921, Apr. 2019.

\bibitem{5629456}
T.~J. {Richardson} and R.~L. {Urbanke}, ``The capacity of low-density
  parity-check codes under message-passing decoding,'' \emph{IEEE Trans. Inf.
  Theory}, vol.~47, no.~2, pp. 599--618, Feb. 2001.

\bibitem{8735878}
G.~{Song}, K.~{Cai}, and J.~{Cheng}, ``Union bound analysis and code design for
  multilevel flash memory channels,'' \emph{IEEE Trans. Commun.}, vol.~67,
  no.~9, pp. 5963--5980, Sept. 2019.

\bibitem{9567703}
P.~{Panteleev} and G.~{Kalachev}, ``Quantum {LDPC} codes with almost linear
  minimum distance,'' \emph{IEEE Trans. Inf. Theory}, vol.~68, no.~1, pp.
  213--229, Jan. 2022.

\bibitem{2222222}
L.~P. {Qian}, Y.~{Wu}, S.~{Zhang}, and Q.~{Chen}, ``Pareto optimal power
  control via bisection searching in wireless networks,'' \emph{IEEE Commun.
  Lett.}, vol.~17, no.~4, pp. 709--712, Apr. 2013.

\bibitem{6403864}
H.~{Xiao}, A.~H. {Banihashemi}, and M.~{Karimi}, ``Error rate estimation of
  low-density parity-check codes decoded by quantized soft-decision iterative
  algorithms,'' \emph{IEEE Trans. Commun.}, vol.~61, no.~2, pp. 474--484, Feb.
  2013.

\bibitem{2666666}
X.~{Fang}, Y.~{Xu}, X.~{Li}, Z.~{Lai}, and W.~K. {Wong}, ``Learning a
  nonnegative sparse graph for linear regression,'' \emph{IEEE Trans. Image
  Process}, vol.~24, no.~9, pp. 2760--2771, Sept. 2015.

\bibitem{2333333}
J.~A. {Snyman}, ``Practical mathematical optimization: An introduction to basic
  optimization theory and classical and new gradient-based algorithms,''
  \emph{Springer}, 2005.

\bibitem{6517051}
Y.~{Fang}, G.~{Bi}, Y.~L. {Guan}, and F.~C.~M. {Lau}, ``A survey on protograph
  {LDPC} codes and their applications,'' \emph{IEEE Commun. Surveys Tuts.},
  vol.~17, no.~4, pp. 1989--2016, Fourth {Q}uarter 2015.

\bibitem{1495850}
J.~{Chen}, A.~Dholakia, E.~Eleftheriou, M.~Fossorier, and X.-Y. {Hu},
  ``Reduced-complexity decoding of {LDPC} codes,'' \emph{IEEE Trans. Commun.},
  vol.~53, no.~8, pp. 1288--1299, Aug. 2005.

\bibitem{8338131}
P.~{Chen}, L.~{Shi}, Y.~{Fang}, G.~F. {Cai}, L.~{Wang}, and G.~R. {Chen}, ``A
  coded {DCSK} modulation system over rayleigh fading channels,'' \emph{IEEE
  Trans. Commun.}, vol.~66, no.~9, pp. 3930--3942, Sept. 2018.

\bibitem{9586047}
Y.~C. {Bu}, Y.~{Fang}, G.~{Zhang}, and J.~{Cheng}, ``Achievable-rate-aware
  retention-error correction for multi-level-cell {NAND} flash memory,''
  \emph{IEEE Trans. Comput.-Aided Des. Integr. Circuits Syst.}, pp. 1--1, 2021.

\bibitem{9600574}
Y.~{Fang}, Y.~{Bu}, P.~{Chen}, F.~C.~M. {Lau}, and S.~A. {Otaibi},
  ``Irregular-mapped protograph {LDPC}-coded modulation: A bandwidth-efficient
  solution for {6G}-enabled mobile networks,'' \emph{IEEE Trans. Intell.
  Transp. Syst.}, pp. 1--14, 2021.

\bibitem{2444444}
H.~{Xiao} and A.~H. {Banihashemi}, ``Improved progressive-edge-growth ({PEG})
  construction of irregular {LDPC} codes,'' \emph{IEEE Commun. Lett.}, vol.~8,
  no.~12, pp. 715--717, Dec. 2004.

\end{thebibliography}

\end{document}